# Empowering Business Transformation - The Positive Impact and Ethical Considerations of Generative AI in Software Product Management
## A Systematic Literature Review

Nishant A. Parikh

nishy.parikh@gmail.com

(Capitol Technology University, Laurel, MD 20708, USA)

**Abstract**

Generative Artificial Intelligence (GAI) has made outstanding strides in recent years, with a good-sized impact on software product management. Drawing on pertinent articles from 2016 to 2023, this systematic literature evaluation reveals generative AI's potential applications, benefits, and constraints in this area. The study shows that technology can assist in idea generation, market research, customer insights, product requirements engineering, and product development. It can help reduce development time and costs through automatic code generation, customer feedback analysis, and more. However, the technology's accuracy, reliability, and ethical consideration persist. Ultimately, generative AI's practical application can significantly improve software product management activities, leading to more efficient use of resources, better product outcomes, and improved end-user experiences.

## 1. Introduction

Generative AI, residing at the cusp of the second and third wave of AI evolution as classified by DARPA (2023), is a testament to the transformative potential of AI. While the first wave comprised rule-based AI systems, and the second wave advanced to machine learning and deep learning applications, the third wave anticipates AI systems capable of contextual adaptation, akin to human understanding. Straddling the second and third phases, generative AI leverages statistical learning to generate novel content such as text, images, or music, often utilizing deep learning frameworks such as Generative Adversarial Networks (GANs). As we progress into this intriguing realm, we unlock vast possibilities for technological transformation.

Generative AI refers to artificial intelligence that can create new content, such as images, text, or music, which did not exist before. It often involves learning from a large amount of input data and then generating something new that reflects the patterns or features observed in that data. In recent years, Artificial Intelligence Generated Content (AIGC) has not only intrigued the computer science community but also captivated societal interest with various products developed by major tech firms, including OpenAI's conversational AI system, ChatGPT, and their creative visual AI, DALL-E2. These models, underpinned by deep generative AI using neural networks, have made significant strides over the past decade (Cao et al., 2023). The most famous form of generative AI is the generative adversarial network (GAN), which consists of two neural networks competing to generate suitable content material. Other generative AI strategies include variational autoencoders (VAE) and transformers, including OpenAI's GPT-3, demonstrating unique competencies in producing human-like textual content (Radford et al., 2021).

Software product management is a crucial discipline that encompasses the activities and responsibilities involved in creating, delivering, and maintaining software products. Following the guidelines set by the International Software Product Management Association (ISPMA), software product managers are equipped with the knowledge and skills to effectively navigate the complex landscape of product strategy, planning, development, and management. By employing a holistic approach, software product management aims to align business objectives with customer needs, optimize product features, prioritize requirements, coordinate cross-functional teams, and oversee the entire product lifecycle. With its emphasis on user-centricity, value creation, and Agile methodologies, ISPMA provides a comprehensive framework that empowers software product managers to drive innovation, foster collaboration, and achieve sustainable success in today's dynamic software industry (ISPMA, 2023).

Software product management is a complicated and challenging discipline that includes various activities: market analysis, product analysis, product strategy, product planning, development, marketing, sales, support, and services. The achievement of software product management relies upon the software product manager's ability to understand the users' needs, market trends, and technological improvements. Generative artificial intelligence (GenAI) has emerged as a promising technology that may help software product managers enhance their decision-making and the quality of their products. This systematic literature evaluation aims to offer an overview of the current state of research on the applications of generative AI in software product management, focusing on the software product manager's activities.

The potential applications of generative AI in software product management are vast, and this article offers an in-depth evaluation of how these applications can transform the software product management field. The following sections will discuss the literature review and the benefits and potential limitations of incorporating generative AI in software product management.

### *Problem Statement*

Generative AI is a revolutionary technology with the potential to modernize software product management by automating tasks, improving efficiency, and enhancing customer experience (Peng et al., 2023; Malik et al., 2022; Siggelkow & Terwiesch, 2023). Growing demand to modernize workflow across industries is expected to drive the demand for generative AI applications among industries. The global market for generative AI is expected to reach $109 billion by 2030 (Grandviewresearch, 2023). As per the McKinsey & Company (2022) report, a 67% average share of respondents reported a revenue increase via AI adoption and 79% reported a cost decrease via AI adoption. The general problem is the need for software product managers to access new and innovative tools that can help them improve their decision-making and profitability and enhance the quality of their products. The specific problem is the lack of awareness of generative AI's potential in software product management. This study aims to bridge this knowledge gap, creating a more comprehensive understanding of how generative AI can be utilized in software product management along with the ethical considerations. It will empower product managers to leverage this technology, potentially leading to the creation of superior products.

## 2. Significance of the Study

The advent and rise of generative AI have dramatically transformed numerous industries, including software product management. Despite the increasing adoption of this technology, there is a lack of comprehensive studies that delve into the full extent of its application, impact, and ethical implications within the realm of product management. This gap is precisely what this study aims to fill.

The significance of this study lies in its potential to contribute a more holistic understanding of the role and value of generative AI in software product management. The findings from this research offer insights that are not only academically significant but also invaluable for business leaders, software product managers, and organizations in the tech industry.

According to Grand View Research, the global market for GenAI is expected to reach $109 billion by 2030 (Grandviewresearch, 2023). With GenAI increasingly becoming a significant part of business strategy and operations, understanding how to best utilize GenAI technologies is critical. This study provides qualitative findings that elucidate the practical benefits of generative AI in software product management, such as increased labor productivity, personalized customer experience, accelerated R&D, and the emergence of new business models. These insights are crucial for leaders who are at the helm of GenAI adoption in their organizations.

By discussing the ethical implications of generative AI, this research offers guidelines that ensure the technology's responsible use, a critical aspect that businesses need to consider in their AI strategies. This research also makes significant contributions to the broader discourse on GenAI and its impact on software product management, bridging the gap between academic research and practical application. By doing so, it expands the body of knowledge in applied research, ultimately paving the way for further exploration and development in the field.

The insights derived from this study can be utilized by businesses and software professionals to enhance their AI strategies, leading to improved organizational efficiency, better product outcomes, and superior customer experiences. Furthermore, these findings can stimulate discussion and policy formulation around the ethical use of AI, contributing to the responsible advancement of the technology.

In essence, this research, by exploring generative AI in software product management, will be of immense value for both academia and industry. Its implications are poised to drive innovation, enhance business processes, and inform ethical considerations in AI deployment.

## 3. Literature Review

This systematic literature review is based on a comprehensive analysis of existing studies related to the applications of generative AI in business and product management. The ISPMA (2023) framework is used to study applications in product management. The International Software Product Management Association (ISPMA) Product Management Framework is a well-known model designed specifically for managing software products. Paajoki (2020) identifies the ISPMA Product Management Framework as the best practice framework to adopt for any organization. The aim of this review is to identify and synthesize the current state of research, focusing on key application areas such as market research, product planning, product documentation, product requirements engineering, product development, UI/UX design, customer insights, and Agile software development. The McKinsey 7-S Framework and Lewin's Change Model frame the review to evaluate how an organization can adopt this technology change, laying the foundation for product managers and business leaders to use the technology effectively and efficiently (Peters & Waterman, 1984; Lewin, 1947). Owen et al. (2013)'s Responsible Innovation framework and General Data Protection Regulation (GDPR, 2023) principles frame the discussion on ethics and privacy.

### 3.1 Relevant Studies in the Software Product Management

#### *3.1.1 Idea Generation*

Karim et al. (2022) researched the application of generative AI for idea generation, brainstorming, and producing research clues. Their work builds on prior research that developed a model for generating medical dialogues related to COVID-19. The authors utilized two GPT3-based models: GPT-NEO-125M and GPT-NEO-1.3B. The larger GPT-NEO-1.3B model consistently generated more coherent and interconnected ideas. Although their study focused on the medical domain, specifically COVID-19, they suggest that the principles derived from their findings could be extended to other scientific or specialized disciplines. The findings show that the larger model generates more coherent text with linked ideas.

#### *3.1.2 Market Research*

Large Language Models (LLMs), such as GPT, are a type of AI created to understand and generate human-like language. Brand et al. (2023)'s research demonstrated that these models can be instrumental in understanding customer preferences. They found that, when treated like a randomly selected customer, GPT exhibits responses that are realistic and consistent with values obtained from existing research. Although they employ traditional market research paradigms to underscore the utility of GPT, LLMs may inspire new market research paradigms unrestricted by the constraints of human subject research. The authors also cautioned that LLMs are known to occasionally "hallucinate" and return incorrect information.

#### *3.1.3 Customer Insights*

Siggelkow and Terwiesch (2023) discussed the transformative potential of large language models such as OpenAI's ChatGPT and Google's Bard in enhancing customer experiences. They highlighted that these AI models can aid in recognizing customer needs by interpreting and integrating data, translating these needs into specific requests, and responding to customers with tailored solutions. They emphasized the importance of focusing on the customer, not just the technology, by addressing specific customer pain points and deploying AI to complement, not replace, a firm's unique capabilities.

Additionally, Siggelkow and Terwiesch (2023) underscored the value of the 'repeat' dimension in customer interactions, where firms learn from each interaction to improve future responses. Generative AI systems excel in this area, creating a positive feedback loop that enhances AI's understanding of the customer for more personalized and effective service. The authors warned against viewing AI as a substitute for human labor, suggesting instead that it should be seen as a tool to enhance a firm's capabilities in unique ways. Ultimately, while generative AI has the potential to revolutionize customer experiences, the authors emphasized the importance of strategic deployment to address customer needs and bolster a firm's distinctive value proposition.

Brynjolfsson et al. (2023) investigated the implementation of a generative AI conversational assistant across 5,179 customer support agents, finding that the AI tool notably increased worker productivity, enhanced customer sentiment, and decreased employee turnover rates. Particularly beneficial for newer and less-skilled workers, the AI tool substantially improved problem resolution and customer satisfaction but did not significantly assist the highest-skilled or more experienced workers. An analysis of the text from agent conversations suggested that AI recommendations guided less-skilled workers to communicate more similarly to high-skilled workers.

### *3.1.4 Product Requirements Engineering*

Malik et al. (2022) focused on the critical process of Requirement Engineering (RE), which is integral to software development and involves defining, documenting, and maintaining software requirements. They noted the significance of Software Requirement Specifications (SRS), key deliverables in the software development life cycle, and the problems arising from ambiguities and conflicts within these documents. To tackle this issue, they proposed a two-phase process for automatic conflict detection in SRS documents that works directly with natural language requirements. The first phase involves transforming software requirements into numeric vectors using transformer-based sentence embeddings, with cosine similarity and ROC curves employed to identify potential conflicts. In the second phase, potential conflicts undergo analysis with general and software-specific Named Entity Recognition (NER), and an overlapping entity ratio is used to determine the final set of conflicts. The authors found an improvement in the F1-score for the OpenCoss and IBM-UAV datasets by 4% and 5%, respectively, underlining its effectiveness.

### *3.1.5 Agile Software Development*

In Agile software development, generative AI can support various activities, such as sprint planning, backlog management, and estimation. For instance, Kim et al. (2021) proposed an AI-driven approach for estimating user story points in Agile projects, leveraging natural language processing techniques to analyze and estimate the effort required for user stories.

Fu and Tantithamthavorn (2022) introduced GPT2SP, an innovative Agile story point estimation approach, utilizing a GPT-2 pre-trained language model and Transformer-based architecture. The proposed method surpasses traditional techniques, such as Planning Poker, Analogy, and expert judgment, and Deep-SE, a deep learning-based approach. Through testing on over 23,000 issues spanning 16 open-source projects, GPT2SP exhibited superior accuracy, outperforming within-project estimates by 34%-57%, cross-project estimates by 39%-49%, and enhancing Deep-SE's performance by 6%-47%. A proof-of-concept tool was also developed to clarify the factors influencing the estimates. A practitioner survey highlighted the challenge of story point estimation. However, it also indicated that an AI-based approach with explanations, such as GPT2SP, was deemed more valuable and trustworthy.

Dam et al. (2019) introduced a framework that adapts and integrates various AI technologies to bolster Agile project management. Although areas in Agile project management are still challenging due to insufficient adequate support, their proposed analytics engine aims to provide decision support on multiple fronts. This includes descriptive analytics, a basic level of analytics that most existing Agile project management tools offer, primarily through data visualization via reports, dashboards, and scorecards.

### *3.1.6 Automated Code Generation*

There have been several types of studies on code generation using generative AI techniques such as Transformer, but Peng et al. (2023) found that GitHub Copilot (which uses generative AI technology for code generation), an AI pair programmer providing context-based code suggestions, significantly boosts productivity. The study revealed that the group using Copilot completed tasks 55.8% faster, marking the first experiment of its kind to provide empirical evidence of the potential of AI tools to enhance human productivity. If extrapolated, a 55.8% productivity increase could lead to substantial economic cost savings and notably influence GDP growth.

Khan and Uddin (2022) examined the application of the Generative Pre-trained Transformer-3 (GPT-3) Codex in automating documentation generation in Software Engineering (SE). Codex demonstrates state-of-the-art performance, outperforming prior models by achieving an average BLEU score of 20.63, even under basic settings. Unlike previous approaches that required task- or language-specific retraining or fine-tuning, Codex operated efficiently with one-shot learning, a process explored by randomly selecting one sample from the corresponding training set.

Park et al. (2023) discussed developing and evaluating the ALSI-Transformer, a transformer-based code comment generation model designed to improve source code comprehension in software development. In response to the increasing need for efficient code commenting as the scale of open-source software grows, the ALSI-Transformer uses a novel method of aggregating multimodal information through Gate Network.

### *3.1.7 UI/UX Design*

The integration of AI technology is revolutionizing the Human-Computer Interaction (HCI) and User Experience (UX) landscapes, redefining user research, design, and evaluation methodologies (Xu, 2023). AI-based solutions enhance the UX quality, with increasing adoption prompted by heightened user awareness of technical innovation.

Houde et al. (2022) explored the objectives, obstacles, and operational practices of teams working on application modernization, as showcased by three projects. The authors discovered that user experience (UX) modernization is a complex, labor-intensive segment of the broader modernization process, involving numerous manual tasks performed by multiple team members, such as project managers, UX designers, and software engineers. Despite the crucial role of UX in these projects, this is often inadequately addressed compared to other tasks such as core code transformation into microservices. However, the authors recognized its potential as a promising avenue for implementing generative AI technologies. By identifying specific pain points in the UX modernization process, they proposed using generative AI models as a possible solution. They also envisioned a future scenario where these models could redefine the UX modernization workflow.

## 3.2 Relevant Studies in Ethics and Privacy

Brand et al. (2023) highlighted the problem of "hallucinations" and returning incorrect information with Large Language Models. It requires consistent human reviews to ensure the accuracy of the AI-generated output.

The rise in deep generative AI models has led to increasingly intricate models. However, their performance relies heavily on quality training data. Notably, these models often exhibit the 'black box' problem, limiting their interpretability and causing potential trust issues. Lastly, handling AI-generated content must address critical social concerns to ensure responsible and beneficial usage for society, emphasizing the importance of trustworthiness and responsibility in the field (Cao et al., 2023).

Generative AI presents legal risks, including potential infringement of intellectual property rights, including unresolved legal questions such as ownership and application of copyright, patent, and trademark laws to AI-generated content. Before leveraging the benefits of generative AI, businesses must understand these risks and devise strategies for self-protection. Essential measures include updating vendor and customer contracts with explicit disclosure about the use of generative AI and clauses safeguarding intellectual property rights. Furthermore, confidentiality provisions should be enhanced to prohibit confidential information in AI tool text prompts (Appel et al., 2023).

Dwivedi et al. (2023) discussed several key challenges that the rise of generative AI, such as ChatGPT, presents to the industry. They highlight serious ethical concerns, as these AI models lack the ability to understand or consider ethical and legal issues. This leads to potential misuse, including the production of deepfakes and disinformation. The inherent 'black box' nature of AI systems poses transparency and explainability issues. Biases can be replicated from the training data, unintentionally leading to misinformation. Legal issues abound, with few guidelines for AI development, unclear copyright boundaries, and dubious ownership of AI-generated content. The authors also foresee potential job losses and increased technology dependency following AI incorporation, stunting personal development and leaving organizations vulnerable if technology fails. Additionally, generative AI systems are limited to combining existing information, offering limited originality. The cultural and personal acceptance of AI could lead to a new form of digital divide. Lastly, the effective use of generative AI requires the ability to design efficient prompts, potentially necessitating widespread new skill training.

### 3.3 Relevant Studies in Business and Management

Korzynski et al. (2023) discussed the impact of generative AI on decision-making in business and management, suggesting that it may address certain limitations in human decision-making processes defined by the bounded rationality model. This model, proposed by Herbert Simon (Simon, 1987), highlights human cognitive limitations, imperfect information, and time constraints as factors affecting our ability to make rational decisions. The use of evolutionary algorithms and generative AI, accelerated by advancements in information technology, could enhance optimization theory and procedural rationality, particularly in customer service. Generative AI systems, such as ChatGPT, can provide structured, logical data to aid decision-makers in filtering and organizing options. The article also notes that demographic factors, such as age and gender, influence the adoption and usage of generative AI in decision-making. Younger users tend to be more technologically optimistic and open to using AI tools, while older users may rely more on their existing knowledge. Gender also plays a role, with men often being more inclined towards innovative technologies.

Korzynski et al. (2023) observed that generative AI, such as ChatGPT, is significantly transforming customer services, prompting a reevaluation of existing theories. With capabilities to interact with customers directly, AI could revolutionize service delivery and customer-organization relations. This has implications for relationship marketing theory, where AI can potentially bolster customer relationships through quick, personalized responses, leading to enhanced customer satisfaction and loyalty. Moreover, generative AI can enhance customer satisfaction and support the presumption concept by allowing consumers to play a more active role in their interactions with organizations. Lastly, in terms of customer experience management theory, generative AI offers new ways of delivering positive customer experiences and engagement. However, the relationship among generative AI services, customer engagement, loyalty, and the creation of new value for organizations requires further research.

## 4. Review Questions

The research questions were categorized using Petticrew and Roberts' (2005) population, intervention, comparison, outcome, and context framework. The *population* is product managers & business leaders working in software organizations. *Intervention* is the Organization & Product Manager's role interfered with generative AI. The *outcome* is the applications and ethical considerations of generative AI in the field of Software Product Management. *Context* is software product companies. The study did not include a *comparison*. The following research questions (see Table 1) were the center of the qualitative study.

**Table 1**

*Research Questions and Motivation*

| Research questions | Motivation |
| --- | --- |
| **1.** What are the applications of generative AI in software product management? | The research investigates the application of generative AI in software product management. It also aims to contribute to a deeper understanding of the role of generative AI in software product management and to identify potential areas for further research and development. The research results may assist software product managers wishing to leverage generative AI to improve their products and services. |
| **2.** What are the ethical challenges presented by Generative AI and what strategies can be employed to mitigate these issues? | Generative AI, with its potential to revolutionize business operations through increased automation, improved efficiency, and rapid decision-making, offers an array of benefits. Yet, as with any transformative technology, it brings with it a raft of ethical implications. These include concerns related to bias, transparency, data privacy, and potential misuse of AI-generated content, which, if left unaddressed, could potentially overshadow generative AI's benefits. This requires a comprehensive exploration of the holistic impact of generative AI on businesses, one that not only delves into its benefits but also grapples with the ethical difficulties arising from its use. By addressing this research question, we aim to shed light on the path businesses must navigate to successfully harness the potential of generative AI while remaining ethically compliant and responsible. |

## 5. Review Methods

A literature review provided the data for this qualitative, descriptive study. Based on current and emerging studies on product management topics, we develop current applications of generative AI in software product management.

### 5.1 Data Source and Search Strategy

The search strategy involved querying significant databases, including IEEE Xplore, ACM Digital Library, Google Scholar, EBSCOhost and ProQuest Central. Keywords such as "generative AI," "software product management," "Idea Generation," "Product Design," "Customer Insights," "project planning," "automated code generation," "UI/UX design," "Ethics," and "customer feedback analysis" were used in different combinations to identify relevant studies.

### 5.2 Study Selection

Each data source contains different inclusion/exclusion criteria. For example, Google Scholar does not have a full-text search or source type selection. It does not include popular magazines, newspapers, or internet articles. IEEE Xplore has full texts, conference proceedings, and IEEE-published standards. Likewise, each data source has different basic or advanced search selection criteria. Table 2 shows each data source's search criteria (inclusion and exclusion). The goal was to collect information for the years between 2016 and 2023, including all sources, full text, and English-language only.

**Table 2**

*Database Sources and Inclusion/Exclusion Criteria*

| No. | Source | Search criteria (inclusion/exclusion) |
|---|---|---|
| 1 | Google Scholar | Inclusion: the years 2016–2023; any type |
| 2 | ACM Digital Library | Inclusion: The ACM Full-Text collection; the years 2016–2023 |
| 3 | EBSCOhost | Inclusion: Full text; peer-reviewed, the years 2016–2023<br>Exclusion: Magazines, trade publications |
| 4 | IEEE Xplore | Inclusion: the years 2016–2023 |
| 5 | ProQuest Central | Inclusion: Full text, peer-reviewed, the years 2016–2023<br>Exclusion: Source type: blogs, podcasts, & websites; trade journals, wire feeds, newspapers, magazines |

### 5.3 Data Extraction and Data Synthesis

#### *5.3.1 Applications of GenAI in Software Product Management*

The ISPMA framework (Figure 1) was used to synthesize the study. Based on the literature review, applications of generative AI in software product management were identified and aligned with ISPMA categories to better understand where generative AI can be used in the software product management framework.

**Figure 1**

*Applications of Generative AI with ISPMA Framework*

| Strategic Management | Product Strategy | Product Planning | Development | Marketing | Sales & Fulfillment | Delivery Service & Support |
|---|---|---|---|---|---|---|
| Corporate Strategy | Positioning & Product Definition | Customer Insight | Product Architecture and Management | Marketing Planning | Sales Planning | Service Planning and Preparation |
| Portfolio Management | Delivery Model & Service Strategy | Product Life Cycle Management | Development Environment Management | Value Communication | Customer Relationship Management | Service Execution |
| Innovation Management | Ecosystem Management | Roadmapping | Development Execution | Product Launches | Operational Sales | Technical Support |
| Resource Management | Sourcing | Release Planning | User Experience Design | Opportunity Management | Operational Fulfillment | Operations |
| Compliance Management | Pricing | Product Requirements Engineering | Detailed Requirements Engineering | Channel Preparation | | |
| Market Analysis | Financial Management | | Quality Management | Operational Marketing | | |
| Product Analysis | Legal and IPR Management | | | | | |
| | Performance & Risk Management | | | | | |

*Note*. The light blue cells show the applications of generative AI in the SPM category.

Although generative AI is a ground-breaking technology, it is still evolving. Its critical use case is a human-like content generation that can be applied to any use case requiring content writing (such as the product or technical documentation) or human-like conversations, such as automated customer support to provide customers with more logical and human-like responses along with the necessary information.

**Market Analysis.** Market analysis is a broad category of SPM, which involves market research, user research, competitive analysis, and many more sub-activities. It is still premature to apply GenAI to all activities. Technology can only help with the existing research-based knowledge. It cannot perform primary research by itself, such as carrying out customer interviews, but it may help analyze customer feedback and reduce Product Manager's time in feedback analysis. Karim et al. (2022) researched the application of generative AI for idea generation, brainstorming, and producing research clues. The research found that GenAI generates coherent and interconnected ideas based on the given knowledge base. Although the study was performed in the medical field, it can be applied to any discipline (Karim et al., 2022).

**Positioning and Product Definition.** This is another area GenAI can assist with as its primary use case is content writing. Based on the given market and product analysis knowledge, it can help create product definitions, including product naming, writing and suggesting product features, etc. Nguyen et al. (2021)'s research shows promising results in writing product descriptions for e-commerce products.

However, it can be applied to any industry. Zhang et al. (2022) carried out similar research in product title generation.

**Customer Insight and Support.** Customer Insight is a critical use-case, as the GenAI technology helps in analyzing customer feedback and learning from that feedback, and in providing better and tailored future responses based on customer needs. GenAI technology has exploded the conversational bot market. According to Statista (2023), the size of the chatbot market is forecast to reach around 1.25 billion US dollars in 2025, a significant increase from the market size in 2016, which stood at 190.8 million US dollars. Brynjolfsson et al. (2023) demonstrated that the implementation of a generative AI conversational assistant enhanced productivity, customer sentiment, and reduced employee turnover in customer support, primarily aiding less experienced workers and boosting overall customer satisfaction.

**Product Requirements Engineering and Detailed Requirements Engineering.** These are the other use cases of GenAI that can help PMs write product documentation, including writing and estimative Agile stories (Kim et al., 2021, Fu and Tantithamthavorn, 2022). The incredible beauty of GenAI lies in Reinforcement Learning with Human Feedback (RLHF), which means that organizations or product managers can feed product-related data to the GenAI tool (such as OpenAI GPT API) for training purposes. The tool can help enhance product knowledge and detailed documentation, including writing Agile stories as the product owner. It can also be used as a product knowledge management system, where anyone from the organization or product team can obtain knowledge using the conversational AI tool. PMs can also train the GenAI tool with competitors' information, including their products, attributes, pricing, website content, etc., information that can be leveraged to generate new ideas for positioning the product. Moreover, it can also be used to generate compelling product titles. Although Zhang et al. (2022) introduced a specific model to generate product titles in e-commerce, it can be used in any field.

**User Experience Design.** Generative AI is transforming User Experience (UX) design in notable ways, providing designers with innovative tools to enhance user engagement and satisfaction. Leveraging the power of machine-learning models such as GANs, generative AI can create numerous design variations, thereby aiding in rapid prototyping and experimentation. Furthermore, it allows for personalized experiences by tailoring interfaces based on user preferences and behaviors. It can also predict and address user needs by understanding and interpreting context, enabling more intuitive and responsive designs. Thus, generative AI paves the way for more user-centered, dynamic, and adaptive UX design (Xu, 2023; Houde et al., 2022).

**Development Execution** is another critical area where GenAI is becoming popular. Automatic code generation was a hot topic among researchers. Figure 2 below shows the maximum number of words used across all studies reviewed. There have been several GenAI models available to generate automatic code. GitHub is one of the most popular sites that provides a GenAI plugin to generate automatic code. Peng et al. (2023) found that the Copilot group completed tasks 55.8% faster.

In a six-week pilot study at Deloitte using 55 developers, most users rated the resulting code's accuracy at 65% or better, with most of the code coming from Codex. The Deloitte experiment found a 20% improvement in code development speed for relevant projects (Davenport & Mittal, 2022).

**Figure 2**

*Word Cloud of the Literature Review*

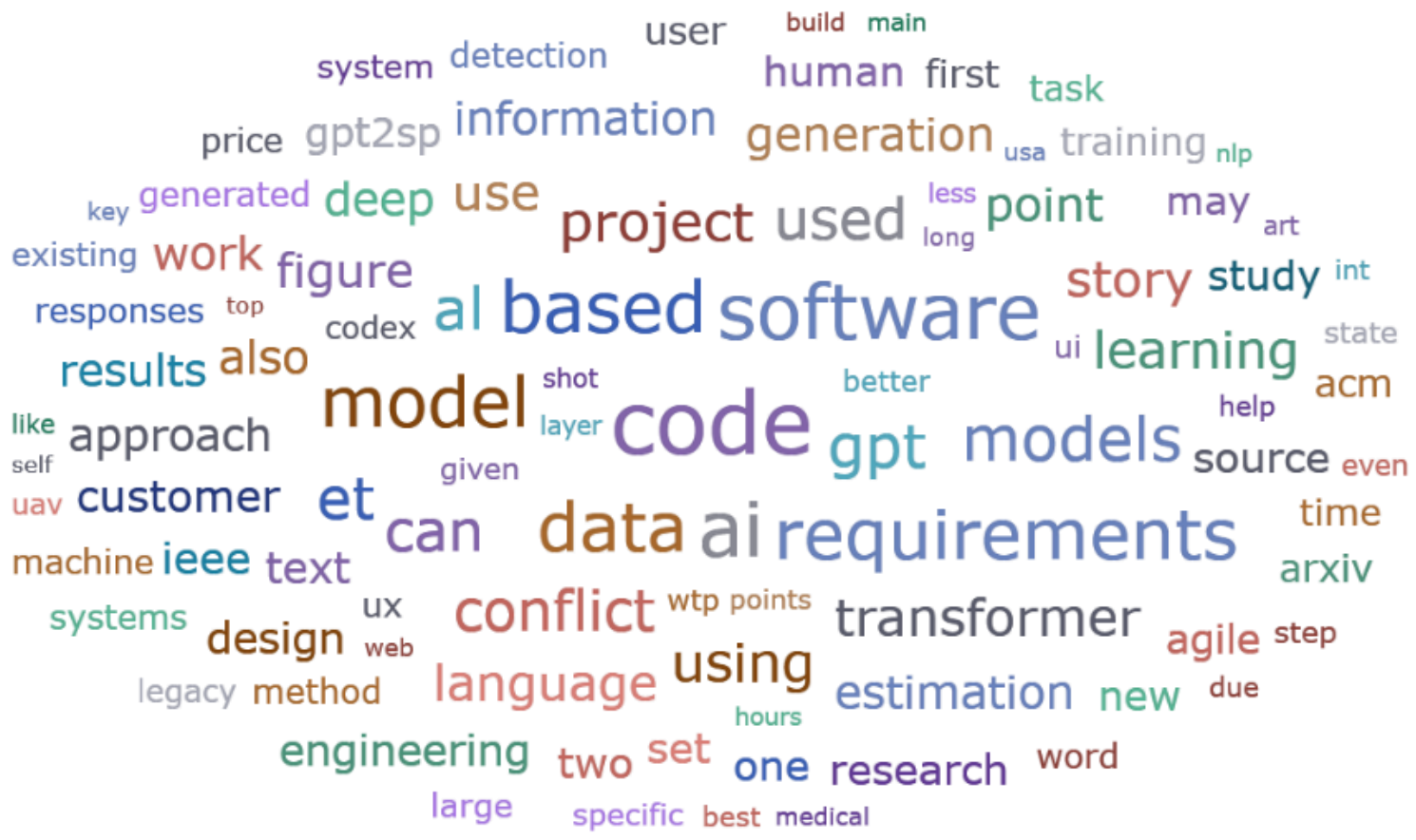

*5.3.2 Change Management and Strategy Models in AI Implementation*

When implementing generative AI, organizations can employ various change management and strategy models. The Lewin Change Model, for example, proposes a three-step process to change: unfreeze, change, and refreeze (Burnes, 2004). This model could be used in the context of introducing generative AI to an organization's existing operations.

Similarly, the McKinsey 7-S Framework emphasizes the importance of aligning seven key organizational elements: strategy, structure, systems, shared values, style, staff, and skills (Peters & Waterman, 1984). This framework could guide the strategic alignment of generative AI with the organization's overall objectives and culture.

**Lewin's Change Model.**

Lewin's (1947) Change Model is a simple yet powerful model for understanding the process of change. The model suggests that change involves a three-stage process: unfreezing, changing (or transition), and refreezing. Here we apply this model to the implementation and acceptance of generative AI in an organization:

***Unfreezing*****:** In this stage, the organization recognizes the need for change. With respect to generative AI, the organization may identify opportunities for increased automation, improved efficiency, reduced manual effort, and improved decision-making (Brynjolfsson et al., 2023; Peng et al., 2023; Fu & Tantithamthavorn, 2022; Khan & Uddin, 2022; Park et al., 2023; Houde et al., 2022; Malik et al., 2022; Korzynski et al., 2023). The organization might also recognize potential limitations, such as concerns about the quality and accuracy of generated outputs, ethical considerations, and difficulties in integrating AI technologies into existing workflows and tools (Brand et al., 2023; Cao et al., 2023; Appel et al., 2023; Dwivedi et al., 2023).

***Changing (Transition)*****:** This is the stage where the actual changes are implemented. In the context of generative AI, this could involve selecting and deploying AI systems, training staff to use them, and making necessary changes to workflows and processes. This stage would require addressing the challenges highlighted in the unfreezing stage. For instance, the organization would need to ensure that the AI systems are transparent and fair and that they can integrate well with existing systems.

***Refreezing*****:** Once the changes are implemented and start to show benefits, the organization goes through a process of reinforcing and stabilizing the change. This could involve setting up procedures for monitoring and evaluating the performance of the generative AI systems, creating guidelines for ethical usage, and making any necessary adjustments to ensure the AI systems continue to deliver benefits while minimizing potential drawbacks.

Using Lewin's Change Model can help organizations understand the process of integrating generative AI, anticipate potential challenges, and devise strategies for successful implementation and ongoing improvement.

**The McKinsey 7-S Framework.**

The McKinsey 7-S Framework is a management model consisting of seven interconnected key elements that are crucial for the success of any organization, and for a company to operate effectively, these elements must be aligned with each other. Here is how we could apply the 7-S framework to better understand the integration of generative AI in an organization:

***Strategy***. The organization should have a clear strategy related to how to leverage generative AI into an existing business model. This could involve automating repetitive tasks, speeding up decision-making processes, or improving workflows. The strategy should also consider potential challenges, such as ethical considerations and the quality and accuracy of outputs.

***Structure***. Incorporating generative AI would necessitate making changes to the organizational structure. Roles and responsibilities may need to be redefined to manage, monitor, and maintain the generative AI systems. For example, DevOps may have to adopt AIOps to manage GenAI applications.

***Systems***. Generative AI can be part of the systems that create, store, and process information. However, careful implementation is necessary to effectively integrate these AI technologies into existing workflows and tools.

***Shared Values***. The use of generative AI should align with the core values of the organization. For instance, if the organization values transparency and accountability, these values should be reflected in the way the AI is designed, trained, and used.

***Skills***. The organization must ensure that its staff has the necessary skills to work with generative AI systems. This could involve training existing staff or hiring new employees with AI expertise.

***Style***. The leadership style within the organization can affect how generative AI is implemented and used. For example, a participative leadership style might involve employees in the decision-making process regarding how and where to use generative AI.

***Staff***. The organization's staff will interact with the generative AI systems. The staff might need to review and ensure the accuracy of the AI-generated outputs. Staff perceptions about the benefits and limitations of generative AI will also affect how readily they adopt these new technologies.

By considering each of these seven elements, an organization can ensure that generative AI is integrated effectively and used in a manner that aligns with its overall mission and values.

### *5.3.3 Ethical Implications of Generative AI*

As generative AI becomes more prevalent, it is essential to consider its ethical implications. Issues such as bias in AI, data privacy, accountability, and the potential misuse of generated content are of concern (Brand et al., 2023; Cao et al., 2023; Appel et al., 2023; Dwivedi et al., 2023). Thus, it is imperative to develop robust ethical guidelines and regulatory measures to ensure the responsible use of generative AI.

Owen et al. (2013)'s Responsible Innovation Framework and General Data Protection Regulation (GDPR, 2023) principles frame the discussion on ethics and privacy.

**Responsible Innovation Framework.** The Responsible Innovation (RI) framework, also known as the AREA framework, was proposed by Owen et al. (2013) to ensure a balanced and ethical approach to the development and application of new technologies. It comprises four dimensions: anticipation, reflexivity, inclusion, and responsiveness.

***Anticipation***. Generative AI has the potential to transform software product management, offering increased automation, improved efficiency, and faster decision-making. By anticipating these benefits, developers and managers can design and implement GenAI strategies that maximize these advantages. However, anticipation must also account for potential issues, including concerns about the quality and accuracy of outputs, ethical considerations such as bias and transparency, and integration challenges. In this regard, potential legal risks, including potential infringement of intellectual property rights, must be carefully considered in advance. This proactive approach can ensure that safeguards are built into the system from the outset, reducing the risk of later issues.

***Reflexivity***. In developing and implementing GenAI, developers must be willing to engage in critical self-reflection. They should scrutinize the decision-making processes surrounding GenAI use, continuously assessing the quality and accuracy of AI-generated outputs. Reflexivity will require human review of AI outputs to rectify any instances of "hallucination" or incorrect information. One should also take the issues related to the 'black box' problem, such as trust and interpretability, into account in this reflection.

***Inclusion***. Generative AI development should involve the input of a diverse range of stakeholders. This can help ensure that the AI system addresses a broad array of needs and concerns, including such ethical considerations as fairness and accountability and such practical issues as integration into existing workflows. Inclusion may also help address any potential legal concerns by involving legal experts who can advise on protecting intellectual property rights and handle any legal issues raised by AI-generated content.

***Responsiveness***. Responsiveness in the context of GenAI would mean being ready to adapt and respond to emerging issues once the technology is implemented. This could involve refining AI algorithms to improve output quality, amending procedures to better integrate AI into existing workflows, or adjusting practices to address ethical issues. On the legal side, responsiveness may involve staying up-to-date with changes in laws and regulations related to AI and intellectual property rights, and updating contracts and policies accordingly.

Applying the RI framework to GenAI can help ensure that this promising technology is used responsibly, with careful consideration given to potential benefits and challenges. It can also help ensure that its use respects ethical principles and legal requirements. This approach can help maximize the benefits of GenAI while minimizing potential risks and harms.

**General Data Protection Regulation (GDPR).** The GDPR was established to protect personal data and privacy rights of individuals within the European Union. It provides comprehensive guidelines and laws related to data collection, storage, and usage. With the rise of generative AI and the substantial data required to train these systems, GDPR's relevance becomes even more critical. One can apply the principles of GDPR to Generative AI, particularly in terms of its benefits and limitations.

The governing principles of the GDPR (2023) as defined in Article 5 are the following:

***Lawfulness, fairness, and transparency (Article 5(1)a).*** GDPR requires that data processing be lawful, fair, and transparent to the data subject. Generative AI systems can enhance transparency by explaining their decision-making process and providing insight into how data is processed. However, the 'black box' nature of these models often makes this challenging, limiting transparency and potentially violating this principle.

***Purpose limitation (Article 5(1)b).*** GDPR stipulates that data collected for specific purposes should not be used for others unless the data subject consents. This aligns with one of the critical ethical considerations in generative AI, ensuring that AI does not deviate from its intended purpose, which can be particularly problematic if it begins generating harmful or discriminatory content.

***Data minimization (Article 5(1)c).*** This principle requires that only the minimum necessary data be collected and processed. Given that generative AI models often require vast amounts of data for training, this can pose a challenge. Businesses must balance the need for data with the requirement to protect individual privacy.

***Accuracy (Article 5(1)d).*** The accuracy of generative AI models is essential in order to avoid the propagation of misinformation or harmful content. This principle aligns with GDPR's requirement for accurate data processing, emphasizing the need for valid and reliable training data.

***Storage limitation (Article 5(1)e).*** GDPR requires that personal data should not be stored longer than necessary. For generative AI systems, this can be complicated, as they may need to retain data for continuous learning and improvement. Businesses must develop strategies to anonymize or delete data once it is no longer needed.

***Integrity and confidentiality (Article 5(1)f).*** Generative AI models, given their reliance on large datasets, must prioritize data security. GDPR mandates that personal data be processed in a way that ensures its security. This includes protecting it from unauthorized access, accidental loss, destruction, or damage.

***Accountability (Article 5(2)).*** Lastly, GDPR requires that organizations take responsibility for all their data processing activities. For generative AI, this could mean auditing and monitoring AI systems to ensure they are functioning as intended and not causing harm or discrimination.

In conclusion, GDPR's principles provide a solid foundation for addressing some of the ethical and practical challenges posed by generative AI. Compliance with these principles can help businesses navigate the complex landscape of AI ethics, protecting individual rights while also leveraging the benefits of this revolutionary technology.

In additional to GDPR, the EU has another AI-specific regulation called the AI Act (The Artificial Intelligence Act, 2023). However, Hacker et al. (2023) argue that the current EU AI Act is ill-equipped to handle the rapid advances and impact of Large Generative AI Models (LGAIMs). They suggest that the current regulatory focus is misdirected, concentrating too much on the direct regulation of AI under the AI Act and neglecting important content moderation issues under the Digital Services Act (DSA). They advocate for more accurate terminology to capture the various actors in the evolving AI value chain, including developers who create and train the models. They caution against technology-specific regulations, which may become rapidly outdated, and suggest a more technology-neutral approach. They stress the urgent need for regulatory updates to maintain online discourse civility, create a level playing field for the development and deployment of new generation AI models, and keep pace with the rapid advancements in AI technologies such as GPT-4.

## 6. Results

In software product management, generative AI has demonstrated potential across various applications, such as market analysis, positioning and product definition, customer insights and support, product requirements engineering, and development execution.

Studies show that large language models, such as GPT-3, can generate more coherent and interconnected ideas, create visually appealing and unique user interfaces, automatically estimate user story points, elaborate on natural language code, and analyze customer feedback (Karim et al., 2022; Kim et al., 2021). Additionally, generative AI can automate the process of requirement elicitation and analysis and software documentation, helping managers better understand and manage product requirements while reducing manual effort. Furthermore, generative AI has the potential to revolutionize market research by using large language models (Brand et al., 2023).

Overall, the applications of generative AI in software product management are vast and diverse, ranging from product discovery to delivery. By leveraging the capabilities of generative AI models, software product managers can automate tedious tasks, make data-driven decisions, and improve overall product quality. The integration of generative AI in software product management processes can potentially lead to better product outcomes, streamlined workflows, and more efficient use of resources across various stages of product development (Karim et al., 2022; Houde et al., 2022; Siggelkow & Terwiesch, 2023; Kim et al., 2021; Fu & Tantithamthavorn, 2022; Dam et al., 2019).

**Research Question 1: What are the applications of generative AI in software product management?**

Table 3 answers the first research question by summarizing the applications of generative AI in software product management.

**Table 3**

*Applications of Generative AI in Software Product Management*

| No. | Application | Summary | Reference |
|---|---|---|---|
| 1 | Market Research | Generative AI models such as GPT can be used in market research, acting as hypothetical customers to understand preferences. Models such as these could inspire new research paradigms unrestricted by human subject research. | Brand et al., 2023 |
| | | Generative AI can generate ideas and brainstorm in specialized disciplines, with larger models generating more coherent and interconnected ideas. | Karim et al., 2022 |
| 2 | Positioning and Product Definition | GenAI, primarily used for content writing, can aid in product positioning and definition, including naming and feature suggestions, with demonstrated efficacy in writing e-commerce product descriptions and title generation across various industries. | Nguyen et al., 2021<br>Zhang et al., 2022 |
| 3 | Customer Insights, | Generative AI models such as ChatGPT and Google's Bard can enhance customer | Siggelkow and Terwiesch, 2023 |

| No. | Application | Summary | Reference |
|---|---|---|---|
| | Customer Support | experiences by interpreting and integrating data, translating needs into requests, and responding with tailored solutions. They can also analyze large volumes of customer feedback data to extract valuable insights. Additionally, generative AI can create a positive feedback loop, enhancing an understanding of the customer for more personalized service. | |
| 4 | Product Requirements Engineering | Generative AI can automate requirement elicitation and analysis, generating natural language requirement statements from structured data. Also, it can help detect conflicts in Software Requirement Specifications (SRS) documents. | Malik et al., 2022 |
| | | Generative AI can support sprint planning, backlog management, and estimation in Agile software development. AI models such as GPT-3 can automatically generate user stories based on provided input data, saving time, reducing human error, and improving the quality of user stories. | Kim et al., 2021; Fu and Tantithamthavorn, 2022 |
| 5 | Development Execution | Generative AI can aid in automated code generation, enhancing productivity and efficiency. Tools such as GitHub Copilot have demonstrated a significant boost in productivity. Furthermore, generative AI can be used to automate documentation generation and improve the comprehension of source code in software development. | Peng et al., 2023; Khan and Uddin, 2022; Park et al., 2023; Houde et al., 2022 |
| | | Generative AI can create visually appealing and unique interfaces, streamlining the design process and enabling rapid prototyping. It can generate UI/UX designs based on user preferences and requirements, leading to more user-centric designs and improved product experiences. | |
| 6 | Decision-making | Generative AI can enhance decision-making in business by addressing limitations of the bounded rationality model, with younger users and males typically more open to this technology's adoption. | Korzynski et al., 2023 |

Figure 3 illustrates the applications of generative AI in SPM.

**Figure 3**

*Applications of Generative AI in Software Product Management*

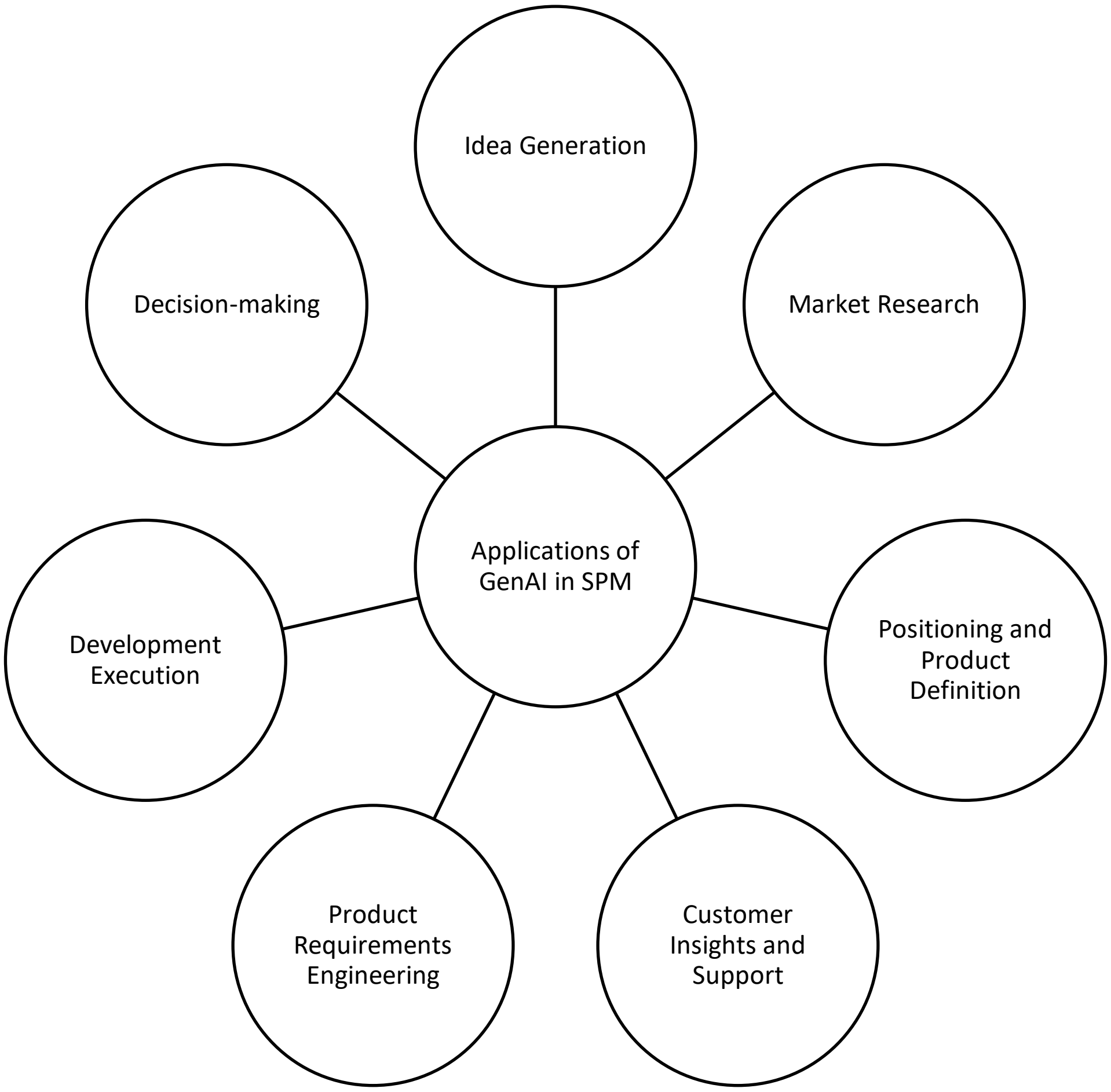

**Research Question 2: What are the ethical challenges presented by Generative AI and what strategies can be employed to mitigate these issues?**

***Ethical Implications***

The ethical implications of generative AI (GenAI) are diverse and consequential, underlining the need for rigorous guidelines and regulatory measures to ensure its responsible usage (Brand et al., 2023; Cao et al., 2023; Appel et al., 2023; Dwivedi et al., 2023).

**Fairness.** Bias is an inherent risk in GenAI. The AI system can inadvertently incorporate and propagate biases present in the data used for training, leading to skewed or prejudiced outputs (Brand et al., 2023). Therefore, it is essential to ensure fairness and diverse training data to avoid potential harm or discrimination.

**Data Privacy** is another critical ethical consideration. As GenAI systems require extensive data for training, the likelihood of infringing on data privacy increases. Adhering to privacy regulations such as the GDPR (2023) is a step towards ensuring data privacy. GDPR's principles provide a comprehensive roadmap for data collection, storage, and usage (GDPR, 2023), guiding businesses in the ethical use of personal data for GenAI.

**Accountability.** Accountability in GenAI involves taking responsibility for the system's outputs. If an AI system generates harmful or inaccurate content, the onus is on the organization to rectify the situation and prevent recurrence (Cao et al., 2023).

**Transparency.** Potential misuse of GenAI-produced content is another pressing concern. There is the risk of fabricated or manipulated content being passed off as legitimate, which could have severe consequences on fields such as news and journalism, and on financial reports and legal documents (Appel et al., 2023). Rigorous verification mechanisms are necessary to prevent such misuse. The increase in deep generative AI models has led to increasingly intricate models. However, their performance relies heavily on quality training data. Notably, these models often exhibit the 'black box' problem, limiting their interpretability and causing potential trust issues (Cao et al., 2023).

**Robustness.** Brand et al. (2023) highlighted the problem of "hallucinations" and returning incorrect information with large language models. This requires consistent human reviews to ensure the accuracy of the AI-generated output.

**Legal Risks.** Generative AI presents legal risks, including potential infringement of intellectual property rights, including unresolved legal questions such as ownership and application of copyright, patent, and trademark laws to AI-generated content. Before leveraging the benefits of generative AI, businesses must understand these risks and devise strategies for self-protection. Essential measures include updating vendor and customer contracts with explicit disclosure about the use of generative AI and clauses safeguarding intellectual property rights. Furthermore, confidentiality provisions should be enhanced to prohibit confidential information in AI tool text prompts (Appel et al., 2023).

The Responsible Innovation (RI) framework, proposed by Owen et al. (2013), offers a comprehensive approach to addressing these ethical issues. The framework encourages anticipation, reflexivity, inclusion, and responsiveness to ensure the balanced development and application of GenAI. The GDPR and RI principles, when appropriately applied, can help mitigate these ethical challenges and guide the responsible use of GenAI.

However, regulatory challenges persist. Hacker et al. (2023) argued that the current EU AI Act fails to keep pace with the rapid advancements and impact of Large Generative AI Models (LGAIMs). The authors advocated for a more technology-neutral approach to regulation, highlighting the need to maintain online discourse civility, create a level playing field for new AI models, and keep up with rapid advancements in AI technologies.

In conclusion, while GenAI promises significant benefits, its ethical implications need careful consideration and proactive management. Using robust ethical guidelines, regulatory measures, and inclusive innovation practices will help maximize the advantages while minimizing potential risks and harms. Figure 4 illustrates ethical implications of generative AI.

**Figure 4**

*Ethical Implications of Generative AI*

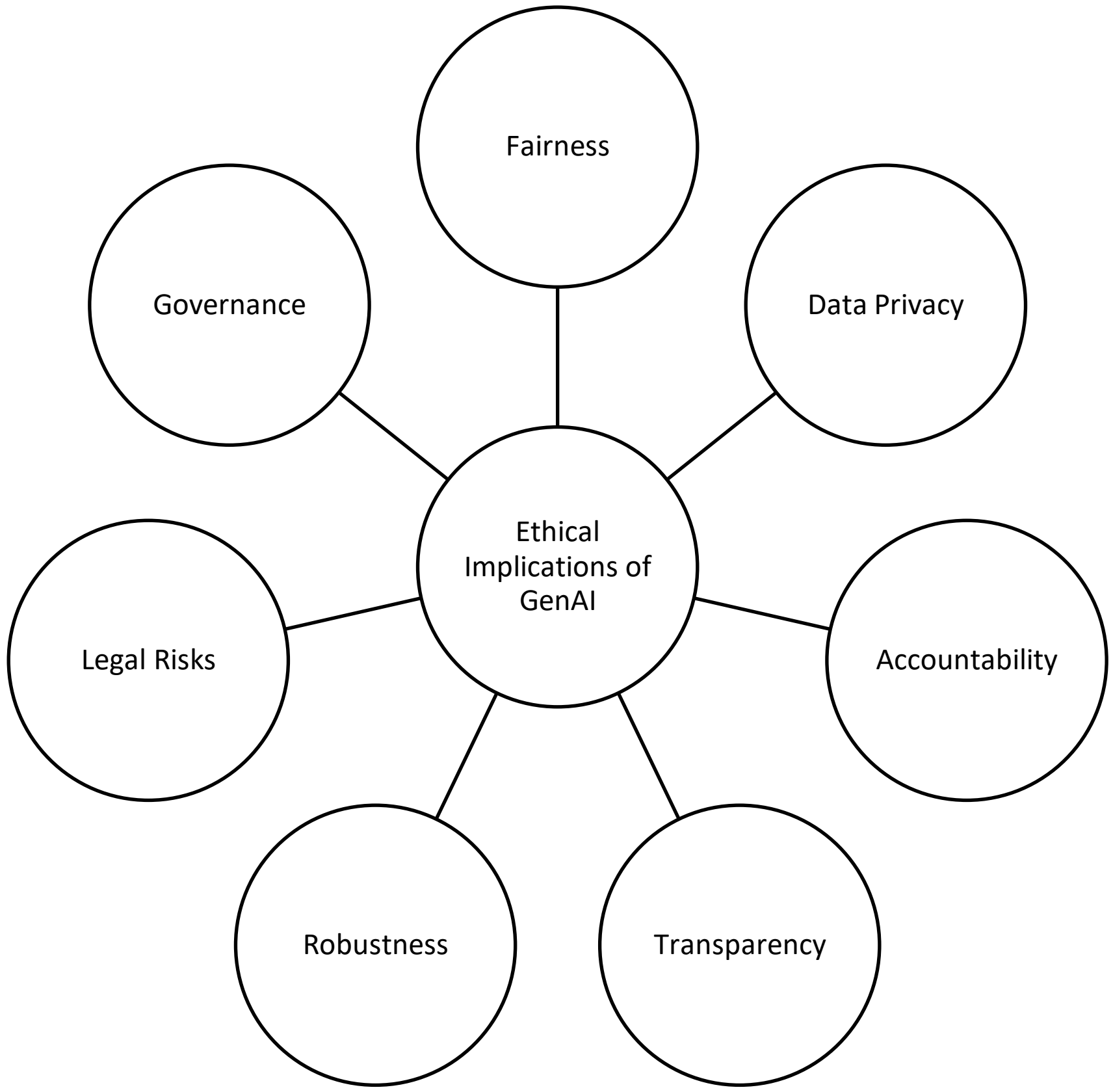

## 7. Conclusions

Generative AI offers promising advancements in business and software product management, helping to transform various aspects such as idea generation, market research, UI/UX design, Agile software development, requirement elicitation, product definition, product development, and customer support. Adopting generative AI technologies, huge language models such as GPT-3 can help software product managers automate time-consuming tasks, make data-driven decisions, and optimize overall product quality.

The role of generative AI in software product management is rapidly evolving, and product managers can use it from ideation to execution. As generative AI technologies evolve, software product managers and researchers must stay abreast of emerging trends and explore innovative ways to use these advancements in software product management effectively.

However, it is essential to acknowledge and address the potential limitations and challenges associated with implementing generative AI solutions in software product management. Bias, transparency, data privacy, accountability, ethical considerations, and integration difficulties should receive careful attention in practice and in future research. Product managers must still review content generated by generative AI to ensure its authenticity and accuracy.

Addressing these ethical concerns requires a proactive and reflexive approach as outlined by the Responsible Innovation Framework (Owen et al., 2013). Anticipating and responding to the challenges of generative AI requires inclusive and responsive strategies that involve diverse stakeholders and adapt to emerging issues. These principles, along with a thorough understanding of legal and regulatory standards, such as the GDPR (2023) and AI Act (2023), can help guide the development and implementation of generative AI technologies (Hacker et al., 2023).

In conclusion, generative AI holds transformative potential for businesses, yet its ethical implications call for comprehensive guidelines, regulatory measures, and a reflexive, anticipatory approach to its development and implementation. It is clear that the future of business will be increasingly intertwined with the responsible and ethical use of generative AI, requiring continuous dialogue, research, and review to ensure its benefits are fully realized while mitigating potential risks and harms.

## 8. Future Research Recommendations

Based on the findings of this systematic literature review, several future research directions can be identified:

Develop more advanced generative AI models: An advanced GenAI model could better understand and cater to the specific needs of software product management processes, such as product strategy, product planning, and feature prioritization.

Integrate AI with Agile methodologies: Exploring how generative AI can be seamlessly integrated with existing Agile methodologies will be crucial for maximizing its potential benefits. It may involve adapting AI-generated user stories, project planning, and resource allocation strategies to align with Agile principles and practices.

Personalize AI assistance for software product managers: As generative AI models become more sophisticated, they can be tailored to individual product managers' needs and preferences. Personalized AI assistants may be able to provide customized recommendations, insights, and support, enabling product managers to work more effectively and efficiently.

Lastly, evaluate the long-term impact of generative AI on software product management workflows, particularly in terms of efficiency, cost-effectiveness, and overall product quality.

The landscape of software product management is poised for significant transformation as generative AI technologies continue to evolve. By staying informed about the latest advances, embracing new applications, and addressing potential limitations and ethical concerns, software product managers can leverage the power of generative AI to drive innovation, efficiency, and success in their projects.

## Acknowledgment

I thank Dr. Burrell (my chair) for insightful feedback and invaluable guidance in reviewing my research. I am also grateful to Capitol Technology University for providing valuable resources on product management. Additionally, I extend my thanks to the anonymous reviewers for their constructive comments. My family, including my spouse and children, have provided constant support throughout this journey, and I am deeply grateful. I would also like to thank my friends for their unwavering motivation. Lastly, I thank all the researchers and authors who have contributed to software product management for their instrumental work in advancing the discipline.